\documentclass[%
aip,
rsi,
 amsmath,amssymb,
 reprint,
superscriptaddress
]{revtex4-2}

\usepackage{graphicx}
\usepackage{dcolumn}
\usepackage{bm}

\usepackage[utf8]{inputenc}
\usepackage[T1]{fontenc}
\usepackage{mathptmx}
\usepackage[export]{adjustbox}
\usepackage{hyperref}
\usepackage{notoccite}
\usepackage{color}
\usepackage{chngcntr}

\makeatletter
\def\@email#1#2{%
 \endgroup
 \patchcmd{\titleblock@produce}
  {\frontmatter@RRAPformat}
  {\frontmatter@RRAPformat{\produce@RRAP{*#1\href{mailto:#2}{#2}}}\frontmatter@RRAPformat}
  {}{}
}%
\makeatother
\begin{document}

\preprint{AIP/123-QED}

\title{A continuous-wave and pulsed X-band electron spin resonance spectrometer operating in ultra-high vacuum for the study of low dimensional spin ensembles}
\author{Franklin H. Cho}
\affiliation{
Center for Quantum Nanoscience, Institute for Basic Science, Seoul 03760, South Korea
}
\affiliation{
Ewha Womans University, Seoul 03760, South Korea
}

\author{Juyoung Park}
\affiliation{
Center for Quantum Nanoscience, Institute for Basic Science, Seoul 03760, South Korea
}
\affiliation{
Department of Physics, Ewha Womans University, Seoul 03760, South Korea
}

\author{Soyoung Oh}
\affiliation{
Center for Quantum Nanoscience, Institute for Basic Science, Seoul 03760, South Korea
}
\affiliation{
Department of Physics, Ewha Womans University, Seoul 03760, South Korea
}

\author{Jisoo Yu}
\affiliation{
Center for Quantum Nanoscience, Institute for Basic Science, Seoul 03760, South Korea
}
\affiliation{
Department of Physics, Ewha Womans University, Seoul 03760, South Korea
}

\author{Yejin Jeong}
\affiliation{
Center for Quantum Nanoscience, Institute for Basic Science, Seoul 03760, South Korea
}
\affiliation{
Department of Physics, Ewha Womans University, Seoul 03760, South Korea
}

\author{Luciano Colazzo}
\affiliation{
Center for Quantum Nanoscience, Institute for Basic Science, Seoul 03760, South Korea
}
\affiliation{
Ewha Womans University, Seoul 03760, South Korea
}

\author{Lukas Spree}
\affiliation{
Center for Quantum Nanoscience, Institute for Basic Science, Seoul 03760, South Korea
}
\affiliation{
Ewha Womans University, Seoul 03760, South Korea
}

\author{Caroline Hommel}
\affiliation{
Center for Quantum Nanoscience, Institute for Basic Science, Seoul 03760, South Korea
}
\affiliation{
Ewha Womans University, Seoul 03760, South Korea
}

\author{Arzhang Ardavan}
\affiliation{
Clarendon Laboratory, Department of Physics, University of Oxford, Oxford OX1 3PU, United Kingdom\looseness=-1
}

\author{Giovanni Boero}
\affiliation{
Microsystems Laboratory, Ecole Polytechnique Fédérale de Lausanne (EPFL), Lausanne 1015, Switzerland\looseness=-1
}

\author{Fabio Donati}
\homepage{donati.fabio@qns.science}
\affiliation{
Center for Quantum Nanoscience, Institute for Basic Science, Seoul 03760, South Korea
}
\affiliation{
Department of Physics, Ewha Womans University, Seoul 03760, South Korea
}


\date{\today}

\begin{abstract}
We report the development of a continuous-wave and pulsed X-band electron spin resonance (ESR) spectrometer for the study of spins on ordered surfaces down to cryogenic temperatures. The spectrometer operates in ultra-high vacuum and utilizes a half-wavelength microstrip line resonator realized using epitaxially grown copper films on single crystal Al$_2$O$_3$ substrates. The one-dimensional microstrip line resonator exhibits a quality factor of more than 200 at room temperature, close to the upper limit determined by radiation losses. The surface characterizations of the copper strip of the resonator by atomic force microscope, low-energy electron diffraction, and scanning tunneling microscope show that the surface is atomically clean, flat, and single crystalline. Measuring the ESR spectrum at 15 K from a few nm thick molecular film of YPc$_2$, we find a continuous-wave ESR sensitivity of $2.6 \cdot 10^{11}~\text{spins}/\text{G} \cdot \text{Hz}^{1/2}$ indicating that a signal-to-noise ratio of $3.9~\text{G} \cdot \text{Hz}^{1/2}$ is expected from a monolayer of YPc$_2$ molecules. Advanced pulsed ESR experimental capabilities including dynamical decoupling and electron-nuclear double resonance are demonstrated using free radicals diluted in a glassy matrix.
\end{abstract}

\maketitle

\section{Introduction}
In recent years, a burgeoning interest has emerged in the investigation of individual magnetic atoms and molecules adsorbed on surfaces, to explore their potential as a new platform for coherently controlled spin systems for quantum operations and quantum sensing.~\cite{Chen23} Such studies of single and coupled atomic and molecular spins on surfaces has predominantly been conducted through a technique that combines electron spin resonance with scanning tunneling microscopy, known as ESR-STM,~\cite{Baumann15,Willke18,Bae18,Veldman21,Yang19,Willke21,Wang23,Wang23science} and more recently, also with atomic force microscopy known as ESR-AFM.~\cite{Sellies23} So far, only few combinations of atoms/molecules and substrates have proven to to be addressable with this technique, and the search for new suitable surface spin systems is often hindered by the complexity and extreme conditions of ESR-STM measurements, typically requiring long experiment times. Moreover, the spins on surfaces have shown relatively short spin coherence times,~\cite{Yang19,Willke21,Wang23,Wang23science} mostly due to the spins experiencing noisy environments from the inherent nature of tunneling current measurements, scattering electrons from the metallic substrates, and magnetic tips of ESR-STM.~\cite{Willke18}

As an alternative to ESR-STM, ensemble-averaging inductive measurements of electron spin resonance (ESR) detecting the absorption and emission of electromagnetic signals by and from ensembles of spins can be used to address surface systems. Ensemble ESR spectroscopy is conventionally conducted on paramagnetic solids (\textit{e.g.}, point defects in crystals/powders), and free radicals in solutions and glassy matrices.~\cite{Weil07} Conversely, very few studies have been performed on ensembles of spins on surfaces as their characterization requires combining additional experimental conditions such as ultra-high vacuum (UHV), metallic substrate in a microwave cavity/resonator, and single crystal surface preparation to a conventional ensemble ESR spectrometer. The earliest experimental efforts of ensemble ESR of spins on surfaces in UHV investigated \textit{in-situ} cleaved silicon surfaces,~\cite{Kaplan1975,Lemke1975} and paramagnetic molecules on metals and inert gases.~\cite{Nilges1980,Nilges1981,Farle1985,Zomack1986,Zomack1987} From the early 90s to mid 2000s, studies of chemisorbed and physisorbed molecules on metals and oxides,~\cite{Katter1993,Schlienz1995,Katter1997}, defects and adatoms on insulators and oxides,~\cite{Schmidt02,Sterrer05,Yulikov06,Benia10,Gonchar11} and hydrogenation and oxidation processes on Si surfaces~\cite{Umeda01,Futako01,Futako04} were conducted using commercial X-band continuous-wave (CW) ESR spectrometers with the additional installations of custom UHV chambers. Most recently, the modification of a commercial W-band CW ESR spectrometer for the study of surface spins was reported.~\cite{Rocker14,Cornu16} Although the spectrometers used in these studies reached the required signal sensitivity in exploring various spins on surfaces, design limitations from utilizing commercial spectrometers, in particular having three-dimensional (3D) metallic cavity resonators and no radiation shield,~\cite{Katter1993,Umeda01,Futako01,Schmidt02,Rocker14} resulted in limited cooling efficiencies (Ref.~\citenum{Katter1993,Schmidt02} and Ref.~\citenum{Rocker14} report the base sample temperatures of 30 K and 50 K, respectively) and distortions of resonator microwave fields by introducing the metallic substrates.~\cite{Katter1993,Schmidt02} Moreover, all of the spectrometers operated only in CW mode and could not probe dynamic properties such as the response to coherent manipulations and decoherence of spins on surfaces.

\begin{figure*}
\includegraphics[width=1.0\textwidth, center]{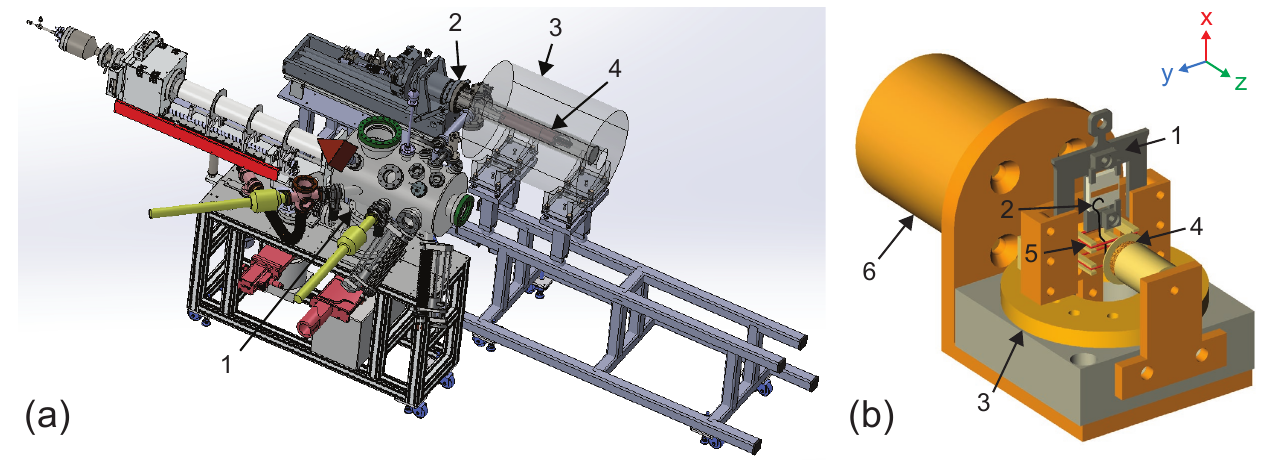}

\caption{Design of the CW/pulsed X-band ESR spectrometer. (a) Overview of the whole instrument including 1. UHV preparation chamber, 2. continuous flow UHV cryostat, 3. 3.2 Tesla cryogen-free superconducting magnet, and 4. radiation shield covering the ESR measurement stage. (b) Detailed drawing of the ESR measurement stage: 1. half-wavelength ($\lambda$/2) microstrip line resonator mounted on a custom flag style sample holder, 2. open-loop antenna inductively coupled to the resonator, 3. piezoelectric rotor (with the rotation axis along the x-axis in the drawing), 4. modulation coil for applying modulating magnetic fields (along the z-direction in the drawing) used in CW ESR experiments, 5. RF coil for applying RF pulses (along the x-direction in the drawing) used in pulsed ENDOR experiments, and 6. cold finger of the cryostat.}
\label{fig1}
\end{figure*}

Here, we describe the development of a custom X-band ESR spectrometer operating in UHV and at cryogenic temperatures which has both CW as well as more sophisticated pulsed ESR experimental capabilities such as dynamical decoupling and electron-nuclear double resonance (ENDOR), aimed at the study of quantum coherence of spins on surfaces. A key feature of our custom spectrometer is that it employs a one-dimensional microstrip line resonator where the Cu strip is epitaxially grown on single crystal sapphire substrate. Our design allows more efficient sample cooling down to 10 K by conduction, and the single crystalline Cu strip also acts as an atomically flat surface on which atomic and molecular spins can be deposited by sublimation. The resonator exhibits a quality factor ($Q$) of more than 200, allowing us to obtain the CW ESR sensitivity of $2.6 \cdot 10^{11}~\text{spins} / \text{G} \cdot \text{Hz}^{1/2}$ at 15 K. This sensitivity corresponds to a signal-to-noise ratio ($SNR$) of $3.9~\text{G} \cdot \text{Hz}^{1/2}$ for an equivalent monolayer of nanometer-size molecular spins, and is one order of magnitude better than previously reported for X-band ESR spectrometers operating in UHV.~\cite{Katter1993,Schmidt02} In addition, the microstrip line allows converting the incident microwave power into the magnetic field with an efficiency that is one order of magnitude higher than the commercial 3D metallic cavity resonators at X-band.

Recent developments in microresonators including those having planar structures have advanced inductive-detection of ESR spectroscopy of mass/volume-limited molecular spin samples.~\cite{Abhyankar22,Abhyankar20,Artzi15,Twig17,Dayan18,Bonizzoni21,Bonizzoni23,Narkowicz05,Torrezan09} These microresonators benefit from higher filling factors for such samples compared to the commerical 3D resonators, thus allowing better ESR sensitivity from small number of spins in samples with limited volume. In particular, the planar geometry of half-wavelength ($\lambda/2$) microstrip line resonators can fulfill the stringent UHV and surface preparation requirements and allow the integration of surface investigation, with guided choice of materials and dimensions. The main goal of this work is to address the complexities and challenges in combining \textit{in-situ} preparation of atomically clean microstrip line resonator suited for surface science studies, integration of the microwave components including our resonator with UHV compatible materials, and cooling principle of the resonator and ESR sample in UHV. Our spectrometer represents a novel tool to investigate the coherent properties of atomic and molecular spins on surfaces and explore low dimensional spin platforms.

This manuscript is constructed as follows: Sect.~\ref{s2ssA} presents the overview of the whole instrument and Sect.~\ref{s2ssB} describes the ESR measurement stage in more detail. The electronics for CW/pulsed ESR experiments are explained in Sect.~\ref{s2ssC} and the microstrip line resonator is characterized by various methods in Sect.~\ref{s3ssA}. Sect.~\ref{s3ssB} estimates the spectrometer's CW ESR sensitivity and Sect.~\ref{s3ssD} demonstrates the spectrometer's pulsed capabilities. Conclusions are provided in Sect.~\ref{s4}. Additional analysis of the resonator quality factor and magnetic field homogenity of the resonator are given in Appx.~\ref{appen_a} and~\ref{appen_b}.


\section{Instrument design}
In this section, we present the detailed design of the spectrometer. Model drawings of the whole instrument and ESR measurement stage are shown, and schematic of electronics for CW/pulsed ESR experiments is explained.

\subsection{UHV preparation chamber, continuous flow UHV cryostat, and 3.2 Tesla cryogen-free superconducting magnet}
\label{s2ssA}

Figure~\ref{fig1}(a) shows an overview of the whole instrument, which includes a UHV preparation chamber, a continuous flow UHV cryostat, and a 3.2 Tesla cryogen-free superconducting magnet. The preparation chamber is equipped with a turbomolecular pump (Pfeiffer Vacuum, HiPace 300), a sputter ion pump (Agilent, VacIon Plus 300), and a titanium sublimation cartridge (Agilent, 916-0050). These components ensure a UHV environment inside the chamber with a base pressure of $<1 \cdot 10^{-9}$ mbar. A load lock with a turbomolecular pump (Pfeiffer Vacuum, HiPace 80) allows introduction of samples without breaking the vacuum in the UHV preparation chamber.

For the preparation of clean metallic surfaces, sublimation of various chemical compounds and characterization of finished surfaces, we have installed a Knudsen effusion cell evaporator with a maximum evaporation temperature of 1200~$^\circ$C (ACME, RHE35B2-M1M210-K-PBN), a 3-cell thermal evaporator with a maximum evaporation temperature of 850~$^\circ$C (Kentax, TCE-BSC), an ion sputter gun (SPECS, IQE 11/35), a thickness/rate monitor with quartz crystal microbalance (Sycon Instruments, STM-100/MF), a custom resistive sample heating station on a long travel (600 mm) XYZ manipulator (custom design from I.T.S) with a differentially pumped rotary platform (MDC Precision, RMTG-450), and high-performance reverse-view low-electron energy diffraction (LEED) optics (SPECS, ErLEED 150). The LEED optics also allow Auger electron spectroscopy (AES) with a maximum energy of 3000 eV. A spring-loaded metal contact is installed on the custom sample heating station to ground the sample surface with a floating potential, preventing charge build-up during sputter cleaning and LEED/AES measurements.

To conduct experiments at cryogenic temperatures, we employ a commercial UHV cryostat (Janis, ST-400) operating with a continuous flow of liquid nitrogen or liquid helium. The cryostat is equipped with a turbomolecular pump (Pfeiffer Vacuum, HiPace 80), a sputter ion pump (Agilent, VacIon Plus 20), and a non-evaporable getter combined ion pump (SAES, NEXTorr\textsuperscript{®} D 200-5) to ensure a UHV enviroment inside the cryostat. When aided by the cryopumping, a base pressure of $<1 \cdot 10^{-10}$ mbar is achieved. The cryostat features an efficient and fast cooldown, typically taking less than 30 minutes to cool its cold finger from room temperature down to 5 K where the consumption of liquid helium is about 4 liters per hour. The cryostat's head hosts mechanical, thermocouple, electrical, and coaxial UHV feedthroughs as well as cryogen entry and vent ports. The CF 100 flange on the neck of the cryostat is bolted onto a differentially pumped rotary platform (McAllister Technical Services, DPRF600), which is mounted on a single-axis linear translator (McAllister Technical Services, ZA6012) with a CF flanged edge-welded bellow having a travel distance of 12 inches. The inner body of the cryostat, mostly enclosed by a gold-coated radiation shield with a sliding cover, moves along with the stretching/compressing motion of the bellow in the linear translator. The radiation shield is normally closed by compression springs, and can be opened by a pull of a linear mechanical feedthrough from the air side. The preparation chamber and cryostat are joined by a 16-inch long full nipple, a gate valve, and a 3-inch long flexible hose (all having CF 40 flange) to account for small misalignments.

As ESR spectroscopy is a magnetic resonance technique requiring an application of an external magnetic field $B_0$ to the sample, we utilize a cryogen-free superconducting magnet (Cryomagnetics, 030-400HTRB-10P-CF) having two solenoid coils made with twisted multi-filamentary NbTi/Cu. The main and sweep coil can apply a maximum field of 3 T and 2000 G with $\sim$70 A and 12 A of current, respectively. The coils and other components inside the magnet are cooled by a pulse tube cryocooler (Sumitomo Heavy Industries, RP-082B2) with a cooling capacity of 1.0 W at 4.2 K which provides the magnet's cryogen-free operation. The magnet sits on a sliding rail system, and its 102-mm diameter room temperature bore allows the magnet to slide over the long outer body of the cryostat which has an outer diameter of 3.75 inches. When slid in fully (with the bellow in the linear translator mounted to the cryostat at its most compressed length), the ESR measurement stage (screwed to the cold finger at the end of the cryostat's inner body) is positioned in the center region of the magnet where the field homogeneity of each coil is less than 10 parts per million within a 1-cm diameter of spherical volume.

\subsection{ESR measurement stage}
\label{s2ssB}

Figure~\ref{fig1}(b) shows a detailed drawing of the ESR measurement stage. A key component in our spectrometer design is an integration of an atomically clean and flat metal single crystal surface and a microwave resonator which opens up a possibility of studying paramagnetic spins on surfaces. Here we employ a half-wavelength ($\lambda$/2) microstrip line resonator made by depositing a strip of copper (approximately 5.1 mm long, 1 mm wide, and 2--3 $\mu$m thick) on a c-plane sapphire substrate (6$\times$6$\times$1 mm) which will be discussed in more detail in the later sections of the paper. The resonator is mounted on a custom flag style sample holder (also known as Omicron plate) with two finger clamps. The use of flag style sample holders allows easy entry and transfer inside the UHV preparation chamber by the load lock door, two storage carousels, two magnetically driven linear/rotary transfer arms, and a wobble stick, all readily available from various manufacturers. To transfer the sample holder between the preparation chamber and the cryostat \textit{in-situ}, the inner body of the cryostat is fully pulled out by 12 inches. This clears the pathway for the 50-inch stroke linear/rotary transfer arm entering from the preparation chamber.

For inductively coupling microwave to the resonator, we use an open-loop antenna made of a thin copper wire. The antenna is connected to the center conductor of a subminiature push-on (SMP) coaxial cable, and the microwave coupling to the resonator can be simply adjusted \textit{in-situ}, even during UHV and cryogenic operations, via pulling out or pushing in the sample holder by a small distance (1--2 mm is more than sufficient) using the wobble stick with a pincer grip installed on the cryostat. This changes the relative positioning of the antenna to the resonator, increasing or decreasing the microwave coupling efficiency.

The resonator mounted on the sample holder inserts into a housing structure which is bolted to a compact piezoelectric rotor (Attocube, ANR240/RES/LT/UHV). The rotor is made with only non-magnetic and low outgassing materials such as BeCu and Ti for UHV and cryogenic operations, and it has a 12.7-mm circular aperture around its rotation axis. The housing structure has a tail part going into the aperture which has a SMP connector made by a glass bead with pin terminal and 2 hole flange mount shroud (the straight end of the antenna described above is soldered to the open end of the bead's pin terminal, which is the opposite side of the SMP connector). As SMP is a push-on or plug-in connector, the connection does not involve any threads to be screwed and thus it can withstand the rotational motions given by the rotor (the center pin of the SMP connector aligns with the rotation axis of the rotor). This allows rotating the normal direction of the resonator surface about $\pm$45$^\circ$ with respect to the axis of the external magnetic field, which can be useful in studying spin systems exhibiting magnetic anisotropies. The limitation in the rotation angle is due to an extrusion of the post for the modulation coil presented in Fig.~\ref{fig1}(b), which allows us to apply modulation fields for lock-in measurements of CW ESR signals. The modulation coil is a multi-turn circular coil, constructed by wrapping Kapton insulated thin copper wire on a post made of polyether ether ketone (PEEK). The direction of the modulation fields generated by the coil is along the same direction as the external field $B_0$. The post for the modulation coil is mounted to a non-rotating part of the rotor, and the direction of the modulation fields is unchanged as the rotor rotates. Moreover, the measurement stage also includes a radio frequency (RF) coil to apply RF pulses for pulsed electron-nuclear double resonance (ENDOR) experiments. The RF coil is a Helmholtz coil, also constructed by wrapping the insulated thin copper wire on another post made of PEEK. The RF fields generated by the coil are perpendicular to $B_0$. The post for the RF coil is mounted to the housing structure and rotates together with the rotor. Since the direction of the RF field is parallel to the rotation axis, it always remains perpendicular to $B_0$ during rotations.

When running an experiment at cryogenic temperatures, all components in the ESR measurement stage are cooled by the conduction from the cold finger. A GaAlAs temperature sensor (Lake Shore Cryogenics, TG-120CU-HT-1.4H) and two Cernox\textsuperscript{®} temperature sensors (Lake Shore Cryogenics, CX-1010-SD-HT and CX-1050-SD-HT-1.4L-P-QL) are installed for monitoring the temperature of the cold finger, the housing structure of the ESR measurement stage, and the radiation shield, respectively. With a continuous flow of liquid helium and pumping of the cryogen vent port, we reach a base temperature of 2.5, 10, and 30 K at the cold finger, the housing structure of the ESR measurement stage, and the radiation shield, respectively.

\begin{figure*}
\includegraphics[width=0.9\textwidth, center]{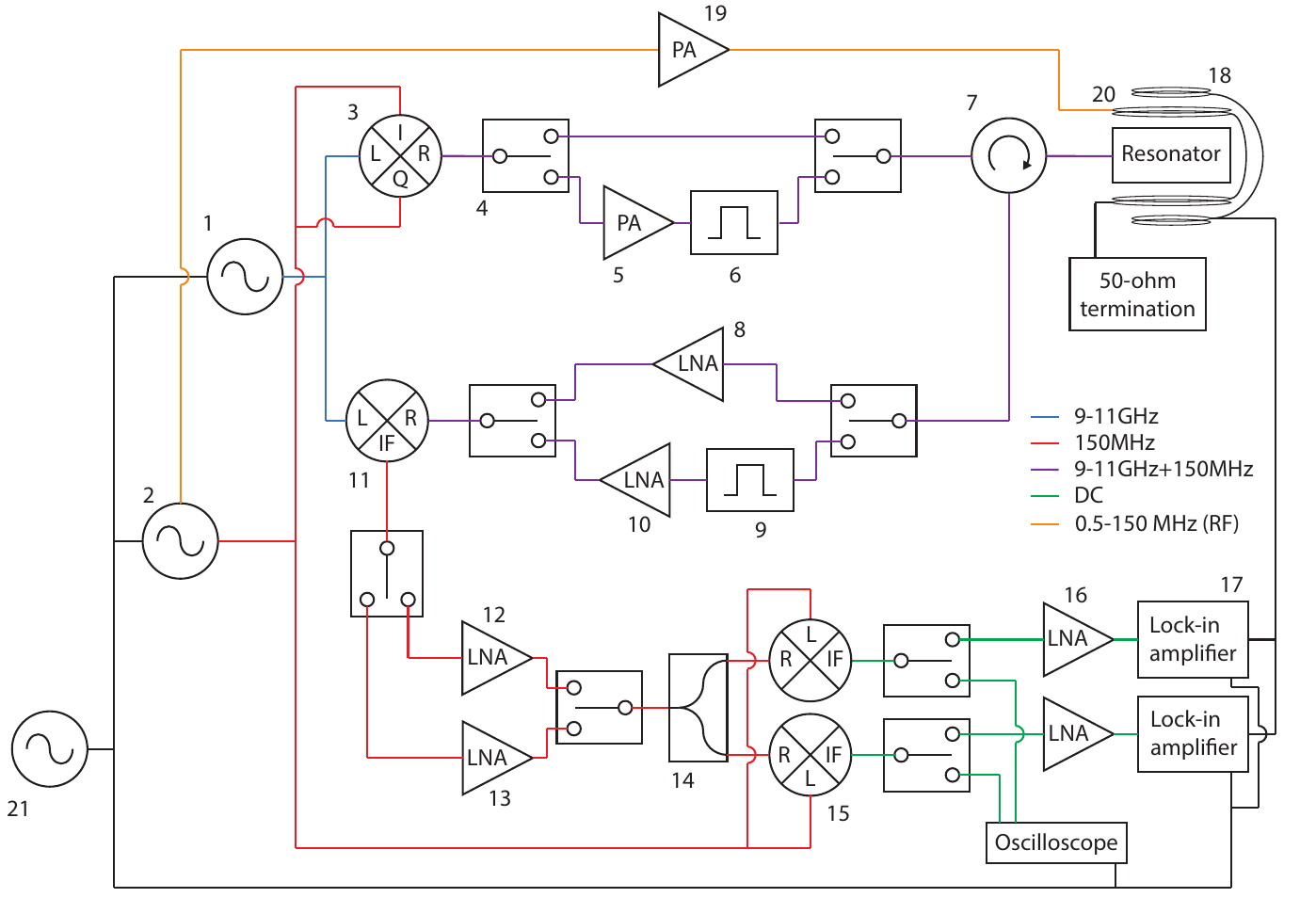}

\caption{Electronics for CW and pulsed ESR experiments. 1. analog signal generator (Keysight Technologies, E8257D), 2. arbitrary waveform generator (Zurich Instrument, HDAWG8), 3. I{\&}Q mixer for single sideband up-conversion (Marki Microwave, MMIQ-0520H), 4. single-pole, double-throw toggle switches (Pasternack, PE7SM1000A; or Fairview Microwave, FMSM1000A), 5. solid-state GaN high power amplifier for pulsed experiments (Qorvo, TGA2238-CP), 6. high-power, fast pin-diode switch to enable/blank pulses/noises from high power amplifier for pulsed experiments (RF-Lambda, RFSP2TR06G13G), 7. circulator (DiTom Microwave, D3C8012), 8. LNA operating in X-band for CW experiments (Qorvo, CMD319C3), 9. fast-pin diode switch for pulsed experiments (Narda Microwave, S213D), 10. LNA operating in X-band for pulsed experiments (RF-Lambda, RLNA08G12G51), 11. first down-conversion mixer to IF (Marki Microwave, M1-0012LQP), 12. LNA operating at IF for CW experiments (Mini-Circuits, ZX60-P103LN+), 13. LNA operating at IF for pulsed experiments (RF-Lambda, RLNA50M05G43), 14. power divider (Mini-Circuits, Z99SC-62-S+), 15. second down-conversion mixers to DC for phase-sensitive, heterodyne detection (Mini-Circuits, ZX05-1HW-S+), 16. pre-amplifiers for CW experiments (Stanford Research Systems, SR560), 17. lock-in amplifiers for CW experiments (Stanford Research Systems, SR860), 18. modulation coil, 19. RF power amplifier (Tomco Technologies, BTM00250-AlphaSA), 20. RF coil, and 21. reference clock (Stanford Research Systems, FS725). For brevity, only the most essential components are shown and components such as various filters, power meters, directional couplers, etc. are omitted.}
\label{fig2}
\end{figure*}

\subsection{Electronics for CW/pulsed ESR experiments}
\label{s2ssC}

Figure~\ref{fig2} shows the electronics for CW and pulsed ESR experiments. We adopt a phase-sensitive heterodyne detection scheme where the excitation microwaves are produced by single sideband up-conversion using an in-phase{\&}quadrature (I{\&}Q) mixer. The local oscillator (LO) signal in 9--11 GHz (X-band) is generated by an analog signal generator. The LO signal is mixed with an intermediate frequency (IF) I{\&}Q signals at 150 MHz provided by an arbitrary waveform generator (AWG). The AWG features the maximum sampling rate of 2.4 GSamples/s, vertical resolution of 16 bits, and 3-dB bandwidth of 750 MHz. For pulsed ESR experiments, the microwave pulses are produced by sending IF pulses instead of CW signals to the I{\&}Q ports. Then, the up-converted pulses after the I{\&}Q mixer are amplified by a power amplifier having the saturation power of 50 W. This is to produce large enough microwave excitation fields $B_1$ at the sample to drive electron spins away from their thermal equilibrium, in a time scale much faster than the spin relaxation times (\textit{e.g.}, it takes about 180 ns to produce a $\pi$ rotation with $B_1$ of 1 G for $g=2$ spins). However, we currently utilize up to 20 W from the amplifier because our high power, fast pin-diode switch, required to blank the amplifier noise for signal detection, has the 1 dB compression point (P\textsubscript{1dB}) of 20 W. Furthermore, the long coaxial connections from the output of the high power switch to the microwave antenna gives an overall loss of about 3 dB at 10 GHz, so the maximum available power incident to the resonator is about 10 W. The amplitudes/phases of the up-converted pulses are easily controlled by adjusting the amplitudes/phases of the IF pulses from the AWG, within the linear regime of the I{\&}Q mixer. Such controls are needed for performing phase cycling and running more sophisticated pulsed experiments.~\cite{Carr1954,Meiboom1958,Schweiger01}

\begin{figure*}
\includegraphics[width=0.95\textwidth, center]{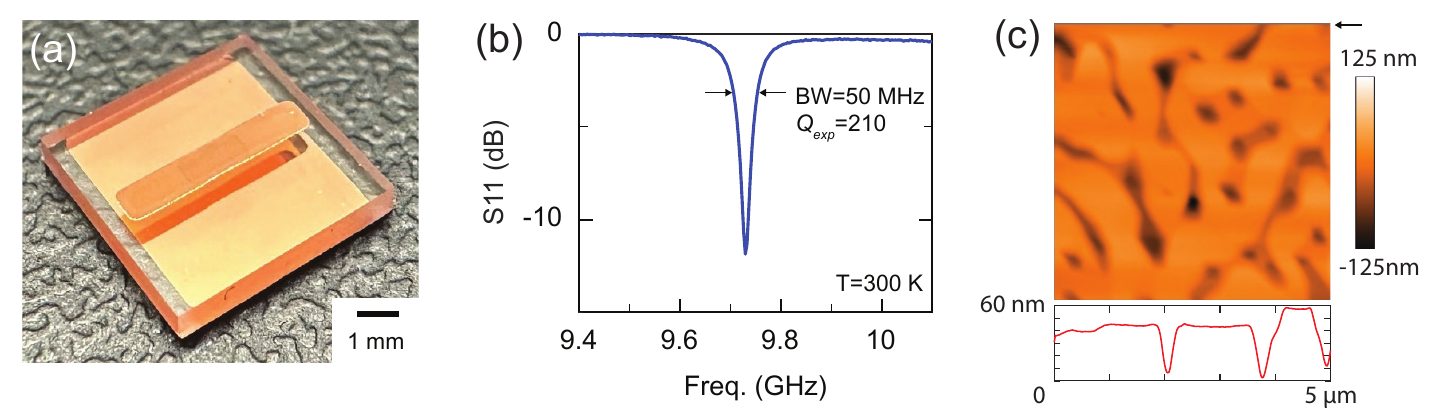}

\caption{Room temperature \textit{ex-situ} characterization of our half-wavelength ($\lambda$/2) microstrip line resonator.  (a) Picture of the resonator. The custom sized cut sapphire substrate is purchased from MTI Corporation. (b) Reflection (S11) measurement showing the resonance from the $\lambda$/2 microstrip line at 9.73 GHz. The observed 3-dB bandwidth (BW) and $Q_{exp}$ of the resonator is 50 MHz and 210, respectively. (c) 5$\times$5 $\mu$m AFM image showing the surface morphology of the copper strip deposited on the c-plane sapphire.  During the 30 minutes of deposition, the substrate temperature was held at 250$\pm$5 $^\circ$C using a thermocouple and a thermal heater monitored and controlled by a custom code written in MATLAB software.~\cite{MATLAB} Color bar on the right indicates the mapping of the height values. Lower trace shows a height profile across the horizontal line along the arrow in the image. The image was taken using a commercial, table-top AFM (Nanosurf AG, NaioAFM) under ambient conditions. The thickness of the strip was measured to be 2.3 $\mu$m using the same AFM (data not shown).}
\label{fig3}
\end{figure*}

The ESR signals, namely, the absorption by the spins for CW ESR experiments or the emission from the spins for pulsed ESR experiments, are guided to the first stage of amplification by a circulator where the amplification is given by a low-noise amplifier (LNA) operating in the X-band. Next, the ESR signals are down-converted to the IF at 150 MHz by mixing with the same LO signals that produced the excitation. Then, the ESR signals at IF are amplified once more by another LNA operating in 50--500 MHz, and down-converted to DC by mixing with two sinusoidal signals at IF. These IF signals having 90$^\circ$ out of phase are also generated by the AWG and allow us to achieve a phase-sensitive detection at the cost of sacrificing 3 dB in the signal power. For CW ESR experiments, a lock-in measurement technique is employed as a final enhancement of the $SNR$. The ESR signals are modulated by applying modulation signals from the lock-in amplifier to the modulation coil which generates small oscillating magnetic fields  on top of the external magnetic field B0 generated by the superconducting magnet. The oscillating fields have a maximum amplitude of $\sim$9 G at 100 kHz, as calibrated by measuring over-modulated linewidths from a sample having an intrinsically narrow linewidth).

For pulsed ESR experiments, the transient responses of the direct emission from spins are captured by a fast digital oscilloscope. Similar to the excitation, the detection phases are controlled by adjusting the phases of the IF sinusoidal signals, enabling phase cycling and suppressing some of unwanted artifacts in capturing pulsed ESR signals.~\cite{Schweiger01} To further reduce overall phase noise in the signals, the analog signal generator, AWG, lock-in amplifiers, and oscilloscope are synchronized to the same reference clock exhibiting ultra-low phase noise characteristics.

For pulsed ENDOR experiments, RF pulses synchronized to the microwave pulses are generated by the AWG and amplified by a RF power amplifier which has the output power rating of 250 W in 0.5--150 MHz. Such amplification is needed to produce large enough RF excitation fields $B_\text{1,RF}$ to drive nuclear spins away from their thermal equilibrium, in a time scale much faster than the spin relaxation times (\textit{e.g.}, it takes about 23 $\mu$s to produce a $\pi$ rotation with $B_\text{1,RF}$ of 5 G for \textsuperscript{1}H spins). Moreover, in order to have a broadband transmission of the RF pulses, the RF coil is terminated with a high power 50-Ohm load in the air side forming a low-pass series RL circuit (resistance of the coil is small compared to 50 Ohm) which is a widely used practice.~\cite{Tschaggelar09,Gromov1999,Reijerse12,Hertel05,Tkach19} Most of the power is dissipated as heat in the 50-Ohm load, but a small fraction of it is used to produce oscillating $B_\text{1,RF}$ fields at the sample by the inductive component of the RF coil.

\section{Instrument performance}
In this section, we investigate the resonance property of the open-circuited $\lambda$/2 line by a reflection (S11) measurement and examine the surface property of the copper strip by atomic force microscope (AFM), low-energy electron diffraction (LEED), and scanning tunneling microscope (STM). Then, we assess the sensitivity of our spectrometer by measuring a CW ESR signal from a magnetically dense, few-layer thick molecular film deposited on top of the copper strip. Finally, we demonstrate various pulsed ESR experimental capabilities of the spectrometer such as dynamical decoupling and ENDOR using free radicals diluted in a bulk, glassy matrix.

\subsection{Characterizations of half-wavelength ($\lambda$/2) microstrip line resonator by reflection (S11), AFM, LEED, and STM}
\label{s3ssA}

Our $\lambda$/2 microstrip line resonator is made by depositing a thick copper strip on top and fully covered bottom surface of a 6$\times$6$\times$1 mm sapphire substrate as shown in Fig.~\ref{fig3}(a). The 6$\times$6 mm surfaces have c-plane (0001) orientation and epitaxy-ready finish with an average surface roughness $R_a$ of $<5$ {\AA} obtained by chemical mechanical polishing. The deposited copper strip are $\sim$5.1 mm long, 1 mm wide, and 2--3 $\mu$m thick. Here, the dimensions are chosen following the general design considerations of a microstrip transmission line and open-circuited $\lambda$/2 resonator.~\cite{Pozar11} In the quasi-static limit where the thickness of the dielectric substrate is much smaller than the wavelength ($\lambda \sim 11$ mm in our case), the effective dielectric constant $\epsilon_e$ of a microstrip line is given by $\epsilon_e = \frac{\epsilon_r + 1}{2} + \frac{\epsilon_r - 1}{2} \frac{1}{\sqrt{1 + 12 d/W}}$ where $\epsilon_r=10$ is the relative dielectric constant of sapphire, $d$ is the thickness of sapphire, and $W$ is the width of the microstrip line. We chose sapphire as our dielectric substrate because of its low dissipation factor, high dielectric constant, and good thermal conductivity. Therefore, copper is a natural choice for metal since it can grow epitaxially on sapphire.~\cite{Katz1968,Dehm1995,Lee14,Verguts16,Kim21} As the beam spot size of our surface characterization tools (LEED/AES) is $\sim$1 mm, we set $W$ to be 1 mm. For a given characteristic impedance $Z_0$, the ratio $W/d$ is found as $W/d=\frac{8e^A}{e^{2A}-2}$ for $W/d<2$ where $A=\frac{Z_0}{60} \sqrt{\frac{\epsilon_r+1}{2}} + \frac{\epsilon_r-1}{\epsilon_r+1} \left( 0.23+\frac{0.11}{\epsilon_r} \right)$.~\cite{Pozar11} Setting $Z_0=50$ Ohm leads to $W/d = 0.96$ or $d = 1.05$ mm, so we went for 1 mm thick sapphire substrate. With $W=d=1$ mm, $\epsilon_e$ is calculated to be 6.75. However, since the calculations are only approximate, several lengths of $l=\lambda/2$ had to be empirically tested on 6$\times$6$\times$1 mm sapphire substrates to obtain the desired resonant frequency around 10 GHz.

Figure~\ref{fig3}(b) shows the reflection (S11) measurement of one of our resonators by a network analyzer (Keysight, N5224B). A sharp resonance at the frequency of 9.73 GHz with resonator quality factor $Q_{exp}=$ 210 is observed. The measurement was carried out while the resonator was mounted on the sample holder, inserted to the housing, and coupled by the antenna as depicted in Fig.~\ref{fig1}(b), with a thin metallic foil (20--40 $\mu$m molybdenum or titanium) over the top side of the resonator (\textit{i.e.}, between the resonator and post for modulation coil; omitted in the drawing for visibility). This provides some shielding for the microwaves at X-band while allowing penetration of modulation fields at 100 kHz. During a cooldown, the tuning undergoes some changes and re-tuning by adjusting the relative position between the resonator and microwave antenna is needed to optimize $Q_{exp}$ again. After re-tuning, the resonator exhibits up to 50\% higher $Q_{exp}$ at base sample temperature. A theoretical analysis of the resonator quality factor reveals that $Q_{exp}$ is dominated by the radiation loss, which is partially improved by the microwave shield (see Appx.~\ref{appen_a}). In addition, finite element method simulations indicate that our resonator offers a vertical homogeneity of $\pm$10\% within 0.1 mm distance from the surface (see Appx.~\ref{appen_b}).

To conduct studies of spins on ordered surfaces, another important feature we must attain is an atomically clean and flat metal single crystal surface of our $\lambda$/2 resonator. To achieve such quality of copper surface having 2--3 $\mu$m thickness on c-plane sapphire, we have employed thermal depositions of pure copper (99.9995\% trace metals basis; from Sigma Aldrich) on the c-plane sapphire surfaces under various deposition, pre-cleaning, and post-treatment conditions.~\cite{Park22} We find good quality copper surfaces appropriate for surface science are realized by depositing copper at fast deposition rates (5--10 {\AA}/s) in high vacuum (HV; $10^{-6}$--$10^{-5}$ mbar) with elevated substrate temperatures (200--350 $^\circ$C), where sapphire substrates are ultrasonicated for 5--30 minutes in isopropyl alcohol and/or acetone before the deposition.~\cite{Park22} The copper depositions are carried out in a separate vacuum chamber with just the sapphire, not mounted to the flag style sample holder. The deposition mask for the microstrip line is put in contact with the sapphire surface to obtain sharp edges on the deposited copper strip (typically a few to couple tens of $\mu$m), and fast deposition rates are achieved by having the evaporator at close distance ($\sim$60 mm) and high temperatures (1000--1200 $^\circ$C). Figure~\ref{fig3}(c) shows an AFM image of such a copper strip surface with no post-treatments where we see a distinctive surface morphology having large flat regions with occasional holes and valleys which are $\sim$30--50 nm deep. The morphology is strongly temperature dependent, and with higher substrate temperature we observe larger flat regions but the holes and valleys get deeper as well.~\cite{Park22}

\begin{figure}
\includegraphics[width=0.5\textwidth, center]{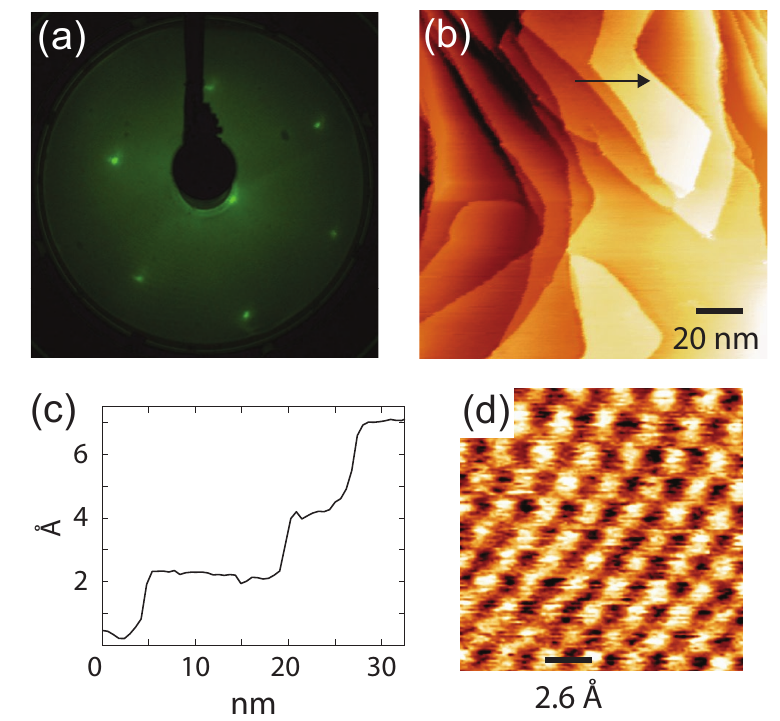}

\caption{Room temperature \textit{in-situ} characterization of our half-wavelength microstrip line resonator. (a) Hexagonal LEED pattern from the copper strip surface. The image was taken at the electron energy of 134 eV after three rounds of 20 minute sputter clean with beam energy of 1 keV and argon pressure of $1 \cdot 10^{-5}$ mbar and subsequent 20 minute anneal at 400 $^{\circ}$C cycle. (b) 150$\times$150 nm STM image taken at room temperature and 1.4 V bias after several rounds of sputter-anneal cycles. The image shows relatively large regions of flat terraces with step edges. (c) Height profile across the horizontal line along the arrow in (b). The profile shows steps corresponding to the mono-atomic layer thickness of Cu(111). (d) 1.55$\times$1.55 nm STM image taken at room temperature and -1 mV bias showing atomically resolved Cu(111) surface.}
\label{fig4}
\end{figure}

To characterize the deposited copper surface using UHV surface characterization tools, we mount the finished microstrip line resonator to the flag style sample holder as Fig.~\ref{fig1}(b) in air and insert it into the preparation chamber via the load lock door. After performing a few cycles of sputtering (20--40 mintues at argon pressure of $10^{-6}$--$10^{-5}$ mbar and ion beam energy of 1--2 keV) and annealing (20--40 minutes around 400 $^\circ$C and $<10^{-7}$ mbar), the Cu surface exhibits good crystallinity as revealed by the sharp spots from the LEED characterization shown in Fig.~\ref{fig4}(a). The hexagonal LEED pattern is a signature of a clean, well-defined Cu(111) surface. We also investigate the same surface using a separate home-built STM where fairly large flat terraces with $\sim$2 {\AA} high steps and atomically resolved Cu(111) surface with atomic spacing of 2.6 {\AA} are observed as shown in Fig.~\ref{fig4}(b), (c), and (d). The LEED and STM measurements show that our deposited copper film surface is atomically clean and ready for surface science. We note that during the sputtering-annealing and LEED measurement carried \textit{in-situ} in our UHV preparation chamber, the floating potential of the copper strip surface is grounded by the spring-loaded metal contact mentioned in Sect.~\ref{s2ssA}. For the STM measurements, the strip is grounded by a fixed piece of metal sheet clamped together to the sample holder.

\begin{figure*}
\includegraphics[width=1.0\textwidth, center]{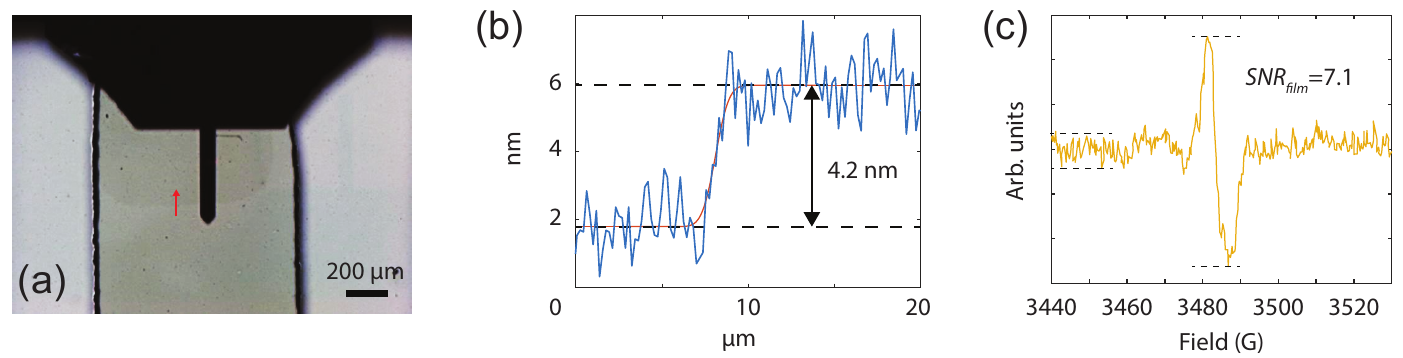}

\caption{CW ESR sensitivity estimation with an amorphous few-layer YPc$_2$ film. (a) Optical image of the film deposited on the center region of the copper strip. A faint contrast is visible as the film is only a few nm thick. Red arrow indicates the scan direction of the subsequent AFM measurement. (b) Height profile of the film measured by AFM under ambient conditions. Blue and red lines are the measurement and fit, respectively. A step of 4.2 nm across the bare copper surface to the deposited YPc$_2$ film is observed. (c) Single-scan CW ESR spectrum of the film measured at 15 K. A single peak centered at 3483 G and having 5 G width is observed with the $SNR_{film}$ of 7.1. For the measurement, a magnetic field sweep rate of 1 G/s and modulation field strength of 4 G are used.}
\label{fig5}
\end{figure*}

We also note that when we deposit copper without additional heating to the substrate (the heat from the evaporator during the deposition heats the substrate slowly to 80 $^\circ$C), the finished surface is not single crystalline (no LEED spots can be observed) but it does not have the few tens of nm deep valleys and holes as those found in the Cu film shown in Fig.~\ref{fig3}(c).~\cite{Park22} There is no noticeable difference in the observed resonator quality factor $Q_{exp}$ between the resonators with different surface finishes (this also supports our assumption in Sect.~\ref{s3ssA} that the additional conductor loss due to surface roughness of the copper strip is negligible).

Cooling in UHV can only be achieved by conduction, which presented a great obstacle for previous studies as discussed in Ref.~\citenum{Katter1993,Schmidt02,Rocker14}. In the setups discussed in Ref.~\citenum{Katter1993,Schmidt02}, the metallic substrates were cooled via thin (0.3 mm) tungsten wires (the use of the thin wires were probably forced to avoid further microwave field distortions and reduction of $Q$),
while larger contact area could be used in the instrument described in Ref.~\citenum{Rocker14}. However, all previous spectrometers operating in UHV lacked a radiation shield (likely not possible due to design constraints from adapting to preset commercial spectrometers), and the base sample temperature Ref.~\citenum{Katter1993,Schmidt02} and Ref.~\citenum{Rocker14} were only $\sim$30 K and 50--70 K, respectively. Our microstrip line resonator and ESR measurement stage design allows the installation of a radiation shield without hindering the performance of the resonator and modulation coil, and our base sample temperature is 10 K as described in Sect.~\ref{s2ssB}.

\subsection{Estimation of CW ESR sensitivity}
\label{s3ssB}
As we aim for the study of spins adsorbed on surfaces, it is crucial to verify that our spectrometer is sensitive enough to detect the ESR signals from such samples. To estimate the sensitivity of our spectrometer, we measured the CW ESR signal from an amorphous few-layer yttrium bis-phthalocyanine (YPc$_2$) molecular film deposited on one of our microstrip line resonators as shown in Fig.~\ref{fig5}(a). YPc$_2$ is known to exhibit a single isotropic ESR spectrum with $g=2$ which is attributed to the spin-1/2 radical delocalized over the Pc$_2$ ligand (yttrium is diamagnetic).~\cite{Komijani18} YPc$_2$ is particularly interesting for surface science experiments because it is possible to deposit the molecule on suitable substrates using thermal sublimation in UHV.~\cite{Komeda11,Robles12,Barhoumi19} Moreover, the room temperature, free induction decay detected spin nutations from magnetically condensed, non-deuterated single crystal powders of YPc$_2$ have been recently observed.~\cite{Boudalis21} Here, we deposit YPc$_2$ on the center region ($A=2 \times 1~\text{mm}^2$) of our copper strip using a deposition mask. To check the thickness of the deposited film, we measure an AFM height profile which shows a 4.2 nm step from the bare copper surface to the deposited film (see Fig.~\ref{fig5}(b)). For the particular film, the deposition is done for 3 minutes at the crucible temperature of 430~$^\circ$C in HV. Figure~\ref{fig5}(c) shows the single-scan CW ESR spectrum of the 4.2 nm amorphous YPc$_2$ film on the microstrip line resonator taken at 15 K using our spectrometer. We observe a single isotropic peak centered at 3484 G with the $SNR_{film}$ of 7.1. If we assume the average volume number density ($n_V$) of the amorphous YPc$_2$ film to be approximately 1 molecule/nm$^3$ (crystallized YPc$_2$ molecules are known to have $n_V$ of 0.84 or 0.78 molecule/nm$^3$ in the \textit{P}2$_1$2$_1$2$_1$ or \textit{Pnma} space group, respectively),~\cite{Boudalis21,Komijani18,Katoh09} we can calculate the number of YPc$_2$ molecules or spins in the film as $N_{film}=n_V \cdot V_s = 1~\text{molecule/nm}^3 \cdot \left(2 \times 1~\text{mm}^2 \times 4.2 ~\text{nm}\right) = 8.4 \cdot 10 ^{12}$ molecules or spins. Then, we can estimate the CW ESR sensitivity of our spectrometer as
\begin{align*}
N_{min} = \frac{N_{film}/SNR_{film}}{\Gamma \sqrt{ENBW}} & = \frac{8.4 \cdot 10^{12}~\text{spins} / 7.1}{5~\text{G} \cdot \sqrt{0.83~\text{Hz}}} \\
& = 2.6 \cdot 10^{11}~\text{spins}/\text{G} \cdot \text{Hz}^{1/2}
\end{align*}
where $N_{min}$ is the minimum number of spins required for detection (\textit{i.e.}, $SNR$ of 1) per unit G of linewidth per unit Hz$^{1/2}$ of signal averaging time by the lock-in amplifiers, $\Gamma=5~\text{G}$ is the linewidth of the observed CW ESR peak, and $ENBW=1/4 t_c=0.83~\text{Hz}$ is the equivalent noise bandwidth of the low-pass filter (the filter slope of 6 dB/oct and time constant of $t_c=300~\text{ms}$) used for the lock-in measurement. If we consider a monolayer (ML) of a magnetically dense YPc$_2$ with the average area number density ($n_A$) of approximately 0.5 molecule/nm$^2$ (\textit{i.e.}, one YPc$_2$ molecule in every 2 nm$^2$ area), the number of YPc$_2$ molecules or spins in the 1 ML will be $N_{ML} = n_A \cdot A = 0.5~\text{molecule/nm}^2 \cdot \left( 2 \times 1~\text{mm}^2 \right) = 1 \cdot 10^{12}$ molecules or spins. Then, we can estimate the expected $SNR_{ML}$ from the monolayer of YPc$_2$ molecules as
\begin{align*}
SNR_{ML} & = \frac{SNR_{film} \cdot \Gamma \sqrt{ENBW}}{N_{film}} \cdot N_{ML} \\
& = \frac{7.1 \cdot 5~\text{G} \cdot \sqrt{0.83~\text{Hz}}}{8.4 \cdot 10^{12}~\text{spins}} \cdot 1 \cdot 10^{12}~\text{spins} \\
& = 3.9~\text{G} \cdot \text{Hz}^{1/2}.
\end{align*}
We believe this is a more appropriate way of describing the CW ESR sensitivity in our application, and it means that the $SNR_{ML}$ of 3.9 is expected from 1 ML of molecules having the area number density of 0.5 molecule/nm$^2$ and CW ESR linewidth of 1 G, with $ENBW$ of 1 Hz. In other words, our spectrometer is sensitive down to a coverage of $1/3.9 \sim 0.26$ ML with 1 G linewidth. If the molecules exhibit a wider linewidth and/or line broadening/splitting due to various interactions, the $SNR_{ML}$ would be lower by the increase in the spread of the signal. Lowering $ENBW$ by using a longer $t_c$ and steeper low-pass filter slope (which will increase the signal measurement time and as a result a slower magnetic field sweep rate must be used) and/or averaging multiple scans will allow us to compensate for such decrease in the sensitivity.

We note that the copper strip of the resonator used in this section is deposited while no additional heating was applied to the sapphire substrate as we find this kind of surface is much better suited for measuring a few nm steps from molecule depositions compared to the surface having holes and valleys extending to a few tens of nm as in Fig.~\ref{fig3}(c). In addition, the molecule deposition is done \textit{ex-situ}, \textit{i.e.}, the resonator is brought to air for the in-contact installation of the molecular deposition mask to obtain sharp edges on the deposited molecular film. This was not due to the lack of our instrumental functionalities, but to precisely quantify the amount of YPc$_2$ molecules on the surface of our resonator. The sharp edges, which can be only realized by manually placing a mask in contact with the resonator prior to film deposition, are needed for the precise height profile measurements by AFM as shown in Fig.~\ref{fig5}(b). Sample transfer in vacuum using wobble sticks and transfer arms does not allow precise positioning of the deposition mask in contact with the resonator, which prevents the deposition of YPc$_2$ film having sharp lateral edge and accurate thickness evaluation of the film height by AFM. In addition, it is still unclear whether the first few molecular layers in contact with a clean metal surface would retain their spin properties and allow for ESR detection due to the interaction with the surface conduction electrons. In this respect, the unavoidable of oxide layers on top of the copper surface formed while exposing the resonator in air ensure a natural decoupling of the YPc$_2$ molecules from the conduction electrons of the Cu strip surface, thus allowing the molecules to preserve their pristine spin properties and more reliable estimate of $N_{min}$. Further investigation of the interactions between the bare metal surface and deposited molecules will be a primary objective/goal of future studies.

\subsection{Comparison of CW ESR sensitivity with other ESR spectrometers operating in UHV, and employing $\lambda/2$ microstrip line resonator in non-UHV condition}
\label{s3ssC}

Reference~\citenum{Sterrer05} states that the X-band CW ESR spectrometer operating in UHV~\cite{Katter1993,Schmidt02} has the necessary sensitivity to detect $10^{12}$ spins assuming a typical linewidth for color centers of 1.5 G, and averagings over 12 hours ($ENBW = 1/4t_c = 1/4 \cdot 12~\text{h} = 5.8 \cdot 10^{-6}~\text{Hz}$) at room temperature are performed to achieve their observed $SNR$ from the irradiated defects on MgO surface. Presuming the $SNR$ of 10 is observed with such conditions, $N_{min}$ can be calculated as
\begin{align*}
N_{min} = \frac{N/SNR}{\Gamma \sqrt{ENBW}} & = \frac{10^{12}~\text{spins} / 10}{1.5~\text{G} \cdot \sqrt{5.8 \cdot 10^{-6}~\text{Hz}}} \\
& = 2.8 \cdot 10^{13}~\text{spins} / \text{G} \cdot \text{Hz}^{1/2}.
\end{align*}
Had the measurements been carried at their base temperature of 30 K, the $SNR$ would have been larger by 10 times from the enhanced spin polarization, and $N_{min}$ in that case would be $2.8 \cdot 10^{12}~\text{spins}/\text{G} \cdot \text{Hz}^{1/2}$. The W-band CW ESR spectrometer operating in UHV~\cite{Rocker14} reports CW ESR measurements from $10^{13}$ di-\textit{tert}-butyl nitroxide molecules adsorbed on MgO surface, and states that their sensitivity is comparable with that of the X-band CW ESR spectrometer operating in UHV.~\cite{Katter1993,Schmidt02} From these, we see that the CW ESR sensitivity of our custom spectrometer is an order of magnitude better than that of previously reported X- and W-band spectrometers operating in UHV.~\cite{Katter1993,Schmidt02,Rocker14}

Although the commercial metallic cavity resonator at X-band used in Ref.~\citenum{Katter1993,Schmidt02} and Fabry–Pérot cavity resonator at W-band used in Ref.~\citenum{Rocker14} exhibit higher $Q$ of $\sim$3000 compared to our microstrip line resonator, they are designed to excite a 3D profile (\textit{e.g.}, $\sim$2 or 1 mm in diameter and $\sim$10 or 0.5 mm in height for the volume resonators at X- or W-band, respectively) homogeneously, and are most appropriate for volume samples. Surface spins do not benefit much from such uniformity of the resonator's microwave fields (\textit{i.e.}, these volume resonators with surface spins have extremely low filling factor). Conversely, for microstrip line resonators the excitation volume is mostly localized above the strip as the intensity of the microwave fields rapidly decays moving away from the conductive surface (see Appx.~\ref{appen_b} for more details). For this reason, microstrip line resonators offer much higher filing factors ($\eta = \int_{V_s} B_1^2 \,dV / \int_{V_r} B_1^2 \,dV$ where $V_s$ and $V_r$ are the sample and resonator volume, respectively) for surface spins, and we believe this is a major factor contributing to our superior spin sensitivity compared to previously reported CW ESR spectrometers operating in UHV with volume resonators.~\cite{Katter1993,Schmidt02,Rocker14}

Although the recently developed ESR spectrometers with microresonators~\cite{Abhyankar22,Abhyankar20,Artzi15,Twig17,Dayan18,Bonizzoni21,Bonizzoni23,Narkowicz05,Torrezan09}  are not aimed to operate in UHV with atomically thin samples, the spectrometer described in Ref.~\citenum{Torrezan09} reports the most similar resonator design to ours, namely a microstrip resonator composed of a 2-$\mu$m thick, 1-mm wide, and 6.3-mm long Cu $\lambda/2$ strip line, covered by a 200-nm thick Au, on a 0.68-mm thick alumina substrate exhibiting $Q$ of 60 at the resonance frequency of 8.3 GHz. In their CW ESR sensitivity estimation, they used a well-known 1,1-diphenyl-2picrylhydrazyl (DPPH) radical having a radius of 0.18 mm and height of 0.04 mm ($V_s = 4 \cdot 10^{-12}~\text{m}^3$; $N=7.9 \cdot 10^{15}$ spins with $n_V=2~\text{spins} / \text{nm}^3$). With such geometries, they demonstrated $N_{min}$ of $5.4 \cdot 10^{10}~\text{spins} / \text{G} \cdot \text{Hz}^{1/2}$ at 300 K (which corresponds to $2.7 \cdot 10^9~\text{spins} / \text{G} \cdot \text{Hz}^{1/2}$ at 15 K) where they calculated their $\eta$ to be $1.7 \cdot 10^{-4}$. Although their $N_{min}$ is almost two orders of magnitude lower than what we report in Sect.~\ref{s3ssB}, this is only due to our $\eta$ being much smaller than theirs (our $V_s$ is $2 \times 1~\text{mm}^2 \times 4.2~\text{nm} = 8.4 \cdot 10^{-15}~\text{m}^3$ which is 480 times smaller than their $V_s$). As $N_{min}$ scales inversely proportional to $\eta$ and $Q$,~\cite{Torrezan09,Abhyankar22,Abhyankar20} our $N_{min}$ would have been lower by more than two orders of magnitude (a factor of $4 \cdot 10^{-12}~\text{m}^3 \cdot 60 / 8.4 \cdot 10^{-15}~\text{m}^3 \cdot 210 = 136$ assuming $B_1$ is homogeneous over $V_s$ and the ratio of $\eta$ is given by the ratio of $V_s$). Indeed, our $B_1$ field homogeneity simulations in Appx.~\ref{appen_b} reveal that the $B_1$ given by our resonator is homogeneous ($\pm$10\%) within $2 \times 1 \times 0.1 = 0.2~\text{mm}^3$ or $0.2 \cdot 10^{-9}~\text{m}^3$ volume which is much larger than $V_s$. This means that our $N_{min} $ would have been on the same order of magnitude as $N_{min}$ given in Ref.~\citenum{Torrezan09} if we had made our measurement with same sized sample as Ref.~\citenum{Torrezan09}. In this sense, our sensitivity is limited by the nature of atomically thin samples ($V_s \sim 10^{-15}$ m$^3$ or pL) which we are mainly interested in for the study of spins on surfaces.

\begin{figure*}
\includegraphics[width=1\textwidth, center]{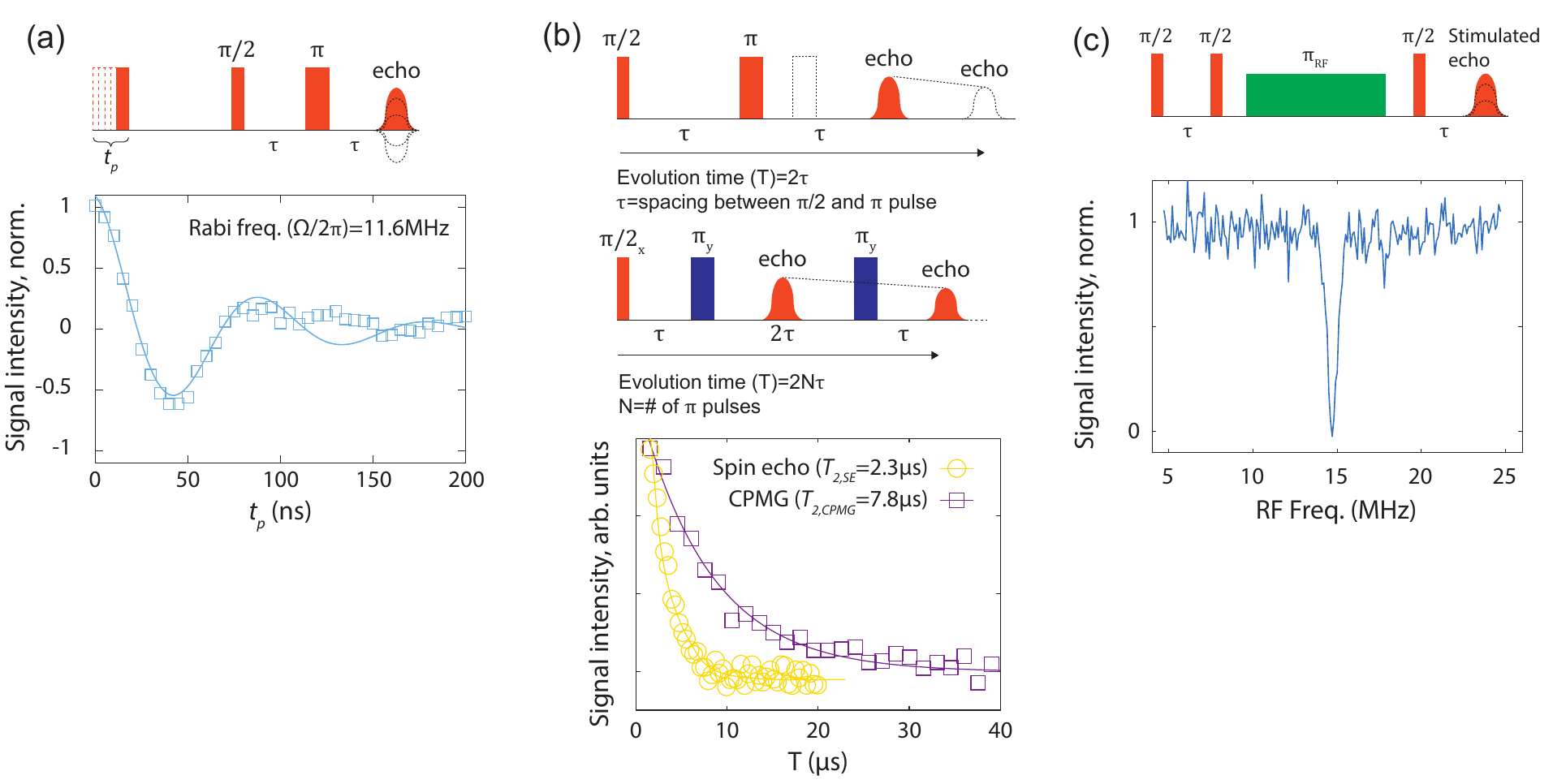}

\caption{Various pulsed ESR experiments with BDPA radicals diluted in PS matrix. (a) Rabi nutation experiment. A $\pi/2$ pulse of 20 ns, a $\pi$ pulse of 40 ns, and a delay time $\tau$ of 700 ns are used. After the nutation pulse of duration $t_p$, a delay of 15 $\mu$s is given before applying the $\pi/2$ pulse. (b) Decay of spin coherence in the transverse plane ($T_2$) measured by SE and CPMG sequence. A $\pi/2$ pulse of 50 ns, a $\pi$ pulse of 100 ns, and a delay time $\tau$ of 700 ns are used. (c) Mims ENDOR experiment. A $\pi/2$ pulse of 50 ns, an RF $\pi$ pulse of 6 $\mu$s (at 250 W), and a delay time $\tau$ of 500 ns are used. All pulsed experiments are performed in ambient conditions, and a repetition rate of 100 Hz and an average of 256 scans (64 $\times$ 4 phase cycling) per data point are used. The spin-lattice or longitudinal relaxation time ($T_1$) is also measured using the inversion recovery sequence, yielding $T_1$ of $\sim$1 ms (data not shown). This validates that the repetition rate of 100 Hz is slow enough (\textit{i.e.}, $>5T_1$)}
\label{fig6}
\end{figure*}

\subsection{Demonstration of various pulsed ESR experimental capabilities}
\label{s3ssD}
Our spectrometer allows performing various pulsed ESR experiments including dynamical decoupling and ENDOR, for potential study of coherence properties of spins adsorbed on surfaces. Currently, only few surface spin systems, \textit{e.g.}, hydrogenated titanium atoms and negatively charged iron phthalocyanine molecules on two ML MgO/Ag(100), have been shown to allow coherent spin manipulation with very short coherence time of few hundreds of ns using ESR-STM.~\cite{Yang19,Willke21} Moreover, MgO films grow in inhomogeneous manner, \textit{i.e.}, one finds a mix of bare Ag(100) and MgO regions with various thicknesses.~\cite{Baumann14} For these reasons, none of the currently known surface spin systems is appropriate for the demonstration of the pulsed ESR experimental capabilities of our spectrometer, and we choose a bulk paramagnetic sample known as 1,3-Bis(diphenylene)-2-phenylallyl (BDPA) radicals diluted in polystyrene (PS) which are widely used in the field of ESR spectroscopy.~\cite{Bennati1999,Rizzato14,Cho14,Cho15,Tkach19,Bonizzoni23} Here we dilute BDPA radicals in PS with 0.1 wt.\% ratio where the total weight of the bulk sample is 2.0 mg. The number of BDPA radicals in the sample can be calculated as $\frac{2.0~\text{mg} \cdot 0.001}{M_{BDPA}} \cdot N_A=2.4 \cdot 10^{15}$ molecules where $M_{BDPA}=495.63$ g/mol is the molar mass of BDPA and $N_A=6.022 \cdot 10^{23}$ molecules/mol is Avogadro's number. Figure~\ref{fig6}(a) shows the Rabi nutation experiment where we observe oscillations of the spin echo (SE) signals from the BDPA:PS sample as a function of the nutation pulse length $t_p$. Fitting the oscillations to a damped sinusoidal function reveals the Rabi frequency ($\Omega /2\pi$) of 11.6 MHz or a $\pi$ rotation of $t_\pi=43$ ns. This means that the $B_1$ fields we generate is $\pi / t_\pi |\gamma_e|=4.2$ G where $|\gamma_e|=1.76 \cdot 10^{11}~\text{rad/s} \cdot \text{T}$ is the electron gyromagnetic ratio. In the experiment, the power of the nutation pulses incident to the sample is held fixed and measured to be 250 mW right after the circulator using a directional coupler and a calibrated power meter in the transmission electronics (omitted in Fig.~\ref{fig2} for brevity). Considering the 3 dB losses from the long coaxial connections from the circulator to the resonator, we provide a pulse power of 125 mW at the resonator. From these, we can calculate how efficient our resonator is in converting the incident microwave power into the magnetic field $B_1$ as $\sim$12 G/W$^{1/2}$ which is an order of magnitude higher than those reported by the commercial 3D metallic cavity resonators at X-band (typically less than a few G/W$^{1/2}$).~\cite{Weber01,Weber02} Again, this enhancement is due to the excitation volume for our microstrip line resonator being much smaller than that of commercial 3D cavities, and was also achieved in other planar and miniature-sized 3D microresonator designs.~\cite{Abhyankar22,Abhyankar20,Artzi15,Twig17,Dayan18,Bonizzoni21,Bonizzoni23,Narkowicz05,Torrezan09}

We note that the size of the bulk BDPA:PS sample we use exceeds the homogeneous volume of our $B_1$ field, especially along the normal direction of the resonator surface as the height ($h$) of the sample is about 1 mm (see Appx.~\ref{appen_b} for more details). This was a compromise we had to make in order to have a reliable weight measurement (\textit{e.g.}, a sample with $h=0.1$ mm would weight 0.2 mg which is too light to be accurately measured with an analytical balance). As a result, we see the non-ideal Rabi oscillations for $t_p$ values longer than 100 ns. However, as the Rabi nutation signals are well described by a damped oscillation function (damping constant is $\sim$50 ns) for $t_p < 100$ ns, spin signals measured using pulses less than 100 ns are coming from the volume having a quite uniform profile of $B_1$. We believe that the $B_1$ generated by our microstrip line resonator is fairly constant up to a certain height $h=h_0$ from its surface, then decays as $1/h$ (see Appx.~\ref{appen_b} for more details). The uncertainty related to the number of spins in the BDPA:PS sample contributing to the SE signals prevents us from obtaining an accurate estimate of pulsed ESR sensitivity of our spectrometer using the observed SNR of pulsed ESR signals as in Ref.~\citenum{Morley08,Hertel05,Cho14}.

Using the same BDPA:PS sample, we also perform SE and Carr-Purcell-Meiboom-Gill (CPMG) decay measurements as shown in Fig.~\ref{fig6}(b). CPMG is one of dynamical decoupling sequences widely used to preserve and enhance the coherence of spins by multiple, equally spaced refocusing $\pi$ pulses having 90$^\circ$ phase difference with respect to the initial $\pi$/2 pulse.~\cite{Carr1954,Meiboom1958,DeLange2010,Bylander11,Cho14,Yoneda18,Jock22,Connors22} Such sequences requiring the control of pulse phases are successfully implemented in our setup by the use of I{\&}Q up-conversion and AWG in the microwave transmission as shown in Fig.~\ref{fig2}. In the CPMG experiment, the coherence of spins is measured as a function of total spin evolution time $\text{T}=\text{2N} \tau$ where N is the number of $\pi$ pulses and $\tau$ is the spacing between the initial $\pi$/2 and first $\pi$ pulse. Here, we add more $\pi$ pulses instead of fixing N and increasing $\tau$ for evolving spins, and we observe that the coherence of spins from the bulk BDPA:PS sample survives for longer T using the CPMG sequence compared to the spin echo sequence. By doing so, the spins at a given $\text{T}$ are refocused more frequently using the CPMG sequence, making them only sensitive to a higher frequency ($f$) noise which is typically lower than a noise at lower $f$ as the noise spectral density of many spin systems follows $1/f$ noise.~\cite{Bylander11,Yoneda18,Jock22,Connors22} Moreover, the 90$^\circ$ phase shift introduced in the $\pi$ pulses compared to the initial $\pi/2$ pulse is known to suppress the systematic pulse angle errors in $\pi$ rotations that otherwise accumulate and cause the reductions of signals as more $\pi$ pulses are applied.~\cite{Meiboom1958,Borneman10,Bylander11} When our signals from the SE and CPMG experiments are fitted to an exponential decay function $e^{-T/T_2}$, we see that the characteristic decay rate $T_{2,CPMG}=7.8~\mu$s is more than 3 times longer than $T_{2,SE}=2.3~\mu$s.

Finally, we demonstrate that we can control not only electron but nuclear spins coherently as well, by measuring the ENDOR signal from the BDPA:PS sample.~\cite{Bennati1999,Rizzato14,Reijerse12,Hertel05,Tkach19} Figure~\ref{fig6}(c) shows the Mims ENDOR spectrum where the electron spin signal from a stimulated echo is monitored while the frequency of the RF pulse which is applied between the second and last $\pi$/2 pulse is varied.~\cite{Schweiger01} When the RF pulse is on-resonance with the $^1$H nuclear spins which are weakly coupled to the electron spins in the BDPA radicals, we observe a reduction of the stimulated echo signal. Here, we are applying a $\pi$ rotation for the $^1$H spins as the signal reduction goes all the way to zero at the nuclear Larmor frequency of $\gamma^{^1\text{H}}_n \cdot B_0=14.7$ MHz where $\gamma^{^1\text{H}}_n=42.58$ MHz/T is the gyromagnetic ratio of $^1$H and $B_0=3450.5$ G is the applied external field.

\section{Conclusion}
\label{s4}
In conclusion, we presented the design and operation of our custom X-band ESR spectrometer optimized for the studies of atomic and molecular spins on surfaces at cryogenic temperatures. The spectrometer features a half-wavelength microstrip line resonator made by depositing $>2~\mu$m thick Cu stripline on the Al$_2$O$_3$ substrate. Reflection measurements show the resonance frequency at X-band having a quality factor $Q_{meas}$ of over 200. The surface characterizations by AFM, LEED, and STM indicate the surface of our copper stripline is atomically clean and flat single crystalline, well-suited for surface science. The continuous-wave ESR measurement of a 4 nm thick amorphous YPc$_2$ film at 15 K reveals that our spectrometer is sensitive enough to detect signals from submonolayer surface spins, and the spin sensitivity of our spectrometer is an order of magnitude better than that of previously reported X-band spectrometers operating in ultra-high vacuum with commercial 3D cavity resonators. Advanced pulsed ESR capabilities including dynamical decoupling and ENDOR are demonstrated using BDPA radicals diluted in a glassy PS matrix.

The instrumentation presented in this paper provides a powerful platform for investigating various interesting spin systems on surfaces such as metal-organic framework complexes~\cite{Willke21} which can be grown in a patterned array,~\cite{Urtizberea18,Noh23} and have been proposed as potential quantum bits for quantum computation and quantum information processing.~\cite{Moreno18,Gaita19,Bayliss20,Carretta21} Moreover, the inductive ESR detection enables measurements of spins which are difficult to read out via electrical transport methods, \textit{e.g.}, lanthanide-based spin systems,~\cite{Gaita19,Reale23,Reale23arXiv} or spins on thick decoupling/insulator layers.~\cite{Paul17,Sellies23} The microstrip line resonator can also be realized with other metal/dielectric and superconductor/dielectric combinations that allow epitaxial growth such as Ag/MgO and Nb/MgO.~\cite{Rodionov19,Fu14}



\begin{acknowledgments}
This work was supported by the Insitute for Basic Science (Grant No. IBS-R027-D1). We thank Andreas J. Heinrich, Junjie Liu, and Alessandro V. Matheoud for heplful discussions. We thank Won-jun Jang for the design and realization of the scanning tunneling microscope, and Soo-Hyon Phark and Hong T. Bui for the support in the surface characterization of the resonator.
\end{acknowledgments}

\section*{Author Declarations}

\subsection*{Conflict of Interest}
The authors have no conflicts to disclose.

\subsection*{Author Contributions}
\textbf{Franklin H. Cho:} Data curation (lead); Formal analysis (lead); Methodology (lead); Software (lead); Visualization (lead); Writing -- original draft (lead). \textbf{Juyoung Park:} Data curation (supporting); Methodology (supporting); Visualization (supporting). \textbf{Soyoung Oh:} Data curation (supporting); Visualization (supporting). \textbf{Jisoo Yu:} Data curation (supporting); Methodology (supporting). \textbf{Yejin Jeong:} Data curation (supporting); Methodology (supporting); Visualization (supporting). \textbf{Luciano Colazzo:} Methodology (supporting); Resources (supporting). \textbf{Lukas Spree:} Resources (equal). \textbf{Caroline Hommel:} Resources (equal). \textbf{Arzhang Ardavan:} Conceptualization (supporting); Writing -- review and editing (supporting). \textbf{Giovanni Boero:} Conceptualization (supporting); Writing -- review and editing (supporting). \textbf{Fabio Donati:} Conceptualization (lead); Formal analysis (supporting); Methodology (supporting); Supervision (lead); Visualization (supporting); Writing -- review and editing (lead).

\section*{Data Availability Statement}
The data that support the findings of this study are available from the corresponding author upon reasonable request.

\appendix
\counterwithin{figure}{section}

\section{Analysis of the resonator quality factor}
\label{appen_a}
For the theoretical consideration of resonator quality factor $Q$, we can write
\begin{align*}
\frac{1}{Q} = \frac{1}{Q_c} + \frac{1}{Q_d} + \frac{1}{Q_r} + \frac{1}{Q_{sw}}
\end{align*}
where $Q_c$, $Q_d$, $Q_r$, and $Q_{sw}$ are contributions from the conductor, dielectric, radiation, and surface-wave losses, respectively.~\cite{Belohoubek1975,Denlinger1980} $Q_c$ in our open-circuited $\lambda$/2 microstrip line resonator can be calculated as $Q_c=\frac{\beta}{2 \alpha_c}=515$ where $\beta=k_0 \sqrt{\epsilon_e}=544$ rad/m is the propagation constant, $\alpha_c=\frac{R_s}{Z_0W}=0.528$ Np/m is the attenuation due to conductor loss, $k_0=\frac{2 \pi f}{c}=209$ m$^{-1}$ is the wavenumber in free space, $R_s=\sqrt{\frac{2 \pi f \mu_0}{2 \rho}}=2.57 \cdot 10^{-2}$ Ohm is the surface resistivity, $\mu_0=4 \pi \cdot 10^{-7}$ H/m is the vacuum permeability, $\rho=5.96 \cdot 10^7$ S/m is the conductivity of copper, and $f=10$ GHz is the microwave frequency.~\cite{Pozar11} $Q_d$ is given by $Q_d=\frac{\beta}{2\alpha_d}$ where $\alpha_d= \frac{k_0 \epsilon_r (\epsilon_e - 1) \tan \delta}{2 \sqrt{\epsilon_e} (\epsilon_r-1)}$ is the attenuation due to dielectric loss.~\cite{Pozar11} As sapphire has low loss tangent $\tan \delta \sim 1 \cdot 10^{-5}$, $\alpha_d$ is two orders of magnitude smaller than $\alpha_c$ and thus $Q_d$ can be neglected in the consideration of $Q$. We note that additional conductor loss can arise from the skin effect and surface roughness of copper strip $R_{a, copper}$, but these become considerable only when the thickness of copper strip and $R_{a, copper}$ are comparable to $\delta_{skin}$.~\cite{Denlinger1980} Our copper strip is more than 3 times thicker and $R_{a, copper}$ is an order of magnitude smaller than $\delta_{skin}=$0.652 $\mu$m at 10 GHz. $Q_r$ in our open-circuited $\lambda$/2 microstrip line resonator is computed as $Q_r=\frac{Z_0}{240 \pi (d/\lambda_0)^2 F(\epsilon_e)}=152$ where $\lambda_0=\frac{c}{f}=0.03$ m is the wavelength in free space and $F(\epsilon_e)=\frac{\epsilon_e+1}{\epsilon_e}-\frac{(\epsilon_e-1)^2}{2 \epsilon_e^{3/2}} \ln \frac{\epsilon_e^{1/2}+1}{\epsilon_e^{1/2}-1}=0.383$ is the radiation form factor.~\cite{Belohoubek1975} Power lost to surface waves is neglected as this becomes dominant loss at much higher frequencies.~\cite{Denlinger1980} We believe the microwave shielding we install reduces the radiation loss resulting in improved $Q_{exp}=210$ compared to the theoretical $Q=\frac{Q_c Q_r}{Q_c+Q_r}=117$. We can estimate the effectiveness of the microwave shield in reducing the theoretical radiation loss by calculating what $Q'_r$ would result in $Q'=Q_{exp}=210$ with $Q_c=515$ as $Q'_r=\frac{Q_c Q'}{Q_c-Q'}=355$, which means reducing theoretical radiation loss by $1-Q_r/Q'_r=67$ \% will result in $Q'=\frac{Q_c Q'_r}{Q_c+Q'_r}=210$.

\section{Finite element simulations of the microwave fields produced by the resonator}
\label{appen_b}

To study the magnetic field profile or $B_1$ field homogeneity of our microstrip line resonator, we performed finite element method simulations using Ansys high-frequency structure simulator (HFSS), a commercial 3D electromagnetic field simulation software for designing and simulating high-frequency electronic products. For the simulations, we modeled a 5.1$\times$1$\times$0.003 mm Cu strip and 6$\times$6$\times$0.003 mm Cu ground plane (bulk conductivity $\rho = 5.8 \cdot 10^7$ S/m) on the top and bottom of a 6$\times$6$\times$1 mm sapphire ($\epsilon_r = 10$) substrate, respectively, in a vacuum ($\epsilon_0 = 1$) volume of 10$\times$10$\times$8 mm around the resonator with radiation absorbing boundaries. The magnitudes of complex magnetic fields $|B_1|$ at the resonance frequency ($\sim$9.85 GHz) were computed using the eigenmode solver of Ansys HFSS.

\begin{figure}
\includegraphics[width=0.5\textwidth, center]{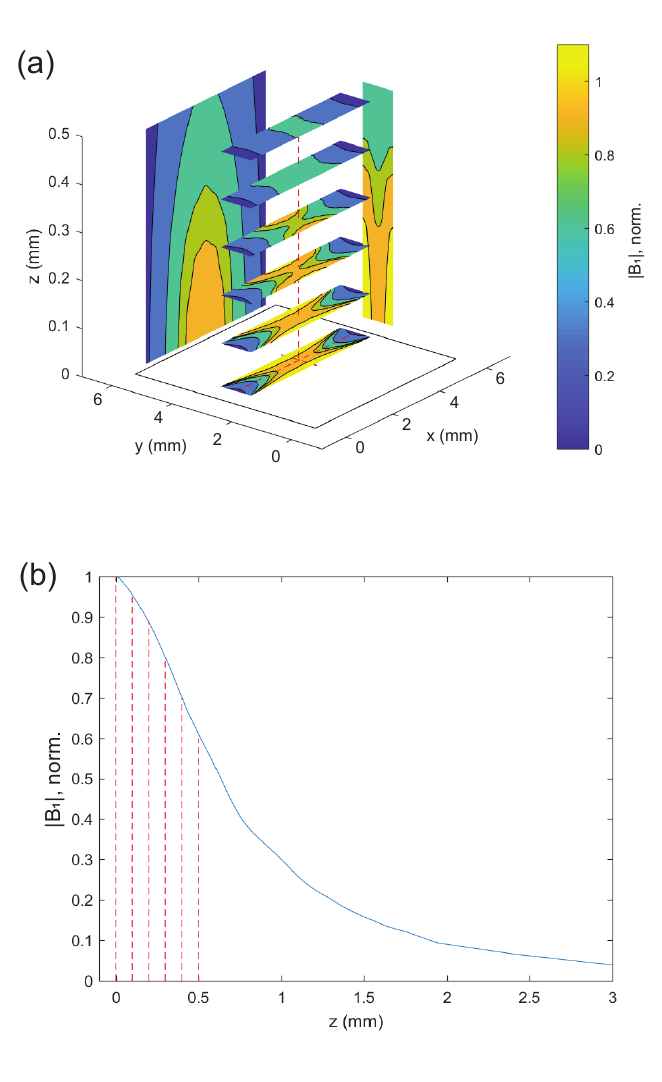}
\caption{Magnetic field profile or $B_1$ field homogeneity on top of the Cu strip at multiple height values. (a) Contour plots of $|B_1|$ field distribution. (b) Decay of $B_1$ as a function of $z$ (vertical distance from the surface of the Cu strip), at $x,y=3$ mm (center of the Cu strip on the $xy$ plane).}
\label{fig_b1}
\end{figure}

Figure~\ref{fig_b1}(a) shows the contour plots of $|B_1|$ on top of the Cu strip (5.1$\times$1 mm) at six height values ($z=0.01$, 0.1, 0.2, 0.3, 0.4, and 0.5 mm where $z=0$ is the interface between the top surface of the sapphire substrate and Cu strip; the Cu strip extends from $z=0$ to 0.003 mm). The values of $|B_1|$ are normalized so that $|B_1|$ at $\left( x,y,z \right) = \left( \text{3, 3, 0.01} \right)$ mm is 1. The contour plots on the $yz$ and $xz$ planes represent $|B_1|$ slicing at $x=3$ and $y=3$ mm, respectively (plotted at shifted positions for visibility). Figure~\ref{fig_b1}(b) shows how $|B_1|$ decays as a function of height $z$ at $x,y = 3$ mm. The simulations reveal that $|B_1|$ within a volume of 2$\times$1$\times$0.1 mm$^3$ or 0.2 $\mu$L on the center of Cu strip is homogeneous, \textit{i.e.},  within $\pm 10$\% of $|B_1|$ at $\left( x,y,z \right) = \left( \text{3, 3, 0.01} \right)$ mm.


\providecommand{\noopsort}[1]{}\providecommand{\singleletter}[1]{#1}%

\end{document}